\documentclass[onecollarge,natbib]{svjour2}
\bibpunct{[}{]}{;}{n}{}{,} % to get "[numbered]" references from natbib
\smartqed  % flush right qed marks, e.g. at end of proof
\usepackage{graphicx}
\usepackage{color}
\usepackage{soul}
\journalname{Few-Body Systems}
\def\Journal#1#2#3#4{{#1} {\bf#2}, #3 (#4)}
\def\NPA{{\rm Nucl. Phys.} A}
\def\NPB{{\rm Nucl. Phys.} B}
\def\PLB{{\rm Phys. Lett.}  B}
\def\PRL{\rm Phys. Rev. Lett.}
\def\PRD{{\rm Phys. Rev.} D}
\def\PRC{{\rm Phys. Rev.} C}

\def\JPG{{\rm J. Phys.} G}
\def\EPJC{{\rm Eur. Phys. J.} C}

\def\ep{\epsilon}

\def\la{\langle}
\def\ra{\rangle}

\def\lam{\lambda}

\def\be{\begin{equation}}
\def\ee{\end{equation}}
\def\bea{\begin{eqnarray}}
\def\eea{\end{eqnarray}}
\begin{document}

\title{Light-Front Quark Model Analysis of Meson-Photon Transition Form Factor
\thanks{This research work is supported by Kyungpook National University Bokhyeon Research Fund, 2015.}
}


\author{Ho-Meoyng Choi         \and
        Chueng-Ryong Ji %etc.
}


\institute{H.-M. Choi \at
              Department of Physics, Teachers Colleges, Kyungpook National University,
              Daegu, 702-701 Korea\\
             \email{homyoung@knu.ac.kr}           %  \\
           \and
           C.-R. Ji \at
           Department of Physics, North Carolina State University, Raleigh, NC 27695-8202, USA \\
           \email{crji@ncsu.edu}
}

\date{Received: date / Accepted: date}

\maketitle

\begin{abstract}
We discuss $(\pi^0,\eta,\eta')\to\gamma^*\gamma$ transition form factors using the light-front quark model.
 Our discussion includes the analysis of the mixing angles for $\eta-\eta'$.
Our results for $Q^2 F_{(\pi^0,\eta,\eta')\to\gamma^*\gamma}(Q^2)$ show scaling behavior for high $Q^2$ consistent with pQCD predictions. 
\keywords{Transition form factor \and $\eta-\eta'$ mixing angle \and Light-front quark model}
\end{abstract}

\section{Introduction}
\label{intro}
The pion-photon transition form factor (TFF) $F_{\pi\gamma}(Q^2)$ has been known to be
the simplest exclusive process involving the strong interaction. It can be calculated
asymptotically at leading twist as a convolution of a perturbative hard scattering amplitude and a gauge-invariant meson distribution
amplitude (DA) which incorporates the nonperturbative dynamics of QCD bound state~\cite{BL80}.
The prediction for $F_{\pi\gamma}(Q^2)$ at the asymptotic limit $Q^2\to\infty$ is shown to satisfy the well-known asymptotic pQCD formula~\cite{BL80}:
$Q^2 F_{\pi\gamma}(Q^2\to\infty)=\sqrt{2}f_{\pi}$ GeV with $f_{\pi}\simeq 130$ MeV.
However, the data for $Q^2F_{\pi\gamma}(Q^2)$ measured from the $\gamma^*\gamma\to\pi^0$ process by the BaBar Collaboration
in 2009~\cite{Babar09} have shown not only the serious violation of the asymptotic pQCD formula but also the rapid growth for $Q^2>15$ GeV$^2$.
On the other hand, in 2012, the Belle Collaboration~\cite{Belle12} has reported their measurement
of the $\gamma^*\gamma\to\pi^0$ process and has shown that the measured values of $Q^2F_{\pi\gamma}(Q^2)$
are consistent with the asymptotic limit of QCD for $Q^2>15$ GeV$^2$.
%These two drastically different experimental results have attracted a lot of attention in hadron
%community and triggered extensive theoretical
%investigations~\cite{MS09,Ra09,MP09,Kroll11,CD10,BCT,MBF}.

Hadronic DAs  provide essential
information on the QCD interaction of quarks, antiquarks and gluons inside the hadrons and
play an essential role in applying QCD to hard exclusive processes.
It has motivated many theoretical studies~\cite{MS09,Ra09,MP09,Kroll11,CD10,BCT,MBF} using various forms of the pion DAs to
understand the discrepancy of $Q^2F_{\pi\gamma}(Q^2)$ data between the BaBar and Belle
measurements. The general agreement on the analysis of the pion DA is that the broader the
pion DA the steeper the slope of $Q^2F_{\pi\gamma}(Q^2)$ as $Q^2$ is getting larger.
For instance,  the flat pion DA~\cite{Ra09,MP09} $\phi(x)=1$ shows the agreement with the BaBar data~\cite{Babar09}.
The subsequent BaBar data~\cite{Babar11} for the $(\eta,\eta')\to\gamma^*\gamma$ TFFs, however,
have shown that the use of flat DA for $\eta$ and $\eta'$ distributions strongly disagrees with the data.
Both Belle data~\cite{Belle12} for $Q^2F_{\pi\gamma}(Q^2)$ and BaBar data~\cite{Babar11} for 
$Q^2F_{(\eta,\eta')\gamma}(Q^2)$
provided consistency with the perturbative QCD prediction and disfavored the flat DA $\phi(x)=1$ which is far different from
the lowest twist-2 DA $\phi(x)=6x(1-x)$ predicted by the asymptotic QCD.
Accordingly, 
%some theoretical efforts have been made for the
%$(\eta,\eta')\to\gamma^*\gamma$ transitions~\cite{DRZ,eta1,eta2,eta3,eta4} as well as $\pi^0\to\gamma^*\gamma$ transition.
careful analysis of  $(\eta,\eta')\to\gamma^*\gamma$ transitions~\cite{DRZ,eta1,eta2,eta3,eta4} appears particularly
important  in the ongoing discussion over the pion-photon TFF results.

The purpose of this work is to comprehensively investigate the $P\to\gamma^*\gamma ~(P=\pi^0,\eta,\eta')$ transitions using the light-front quark model (LFQM)
based on the QCD motivated effective LF Hamiltonian~\cite{CJ_99,CJ_DA}.
The paper is organized as follows. In Sec.~\ref{sec:II}, we discuss the
meson-photon TFFs in an exactly solvable model based on the covariant Bethe-Salpeter (BS) model of
(3+1)-dimensional fermion field theory. Performing both manifestly covariant
calculation and the LF calculation in the BS model, we show the equivalence between the
two results and the absence of the zero-mode contribution to the TTF.
The $\eta-\eta'$ mixing scheme for the calculations of the $(\eta,\eta')\to\gamma^*\gamma$
TFFs is also discussed. We then apply the manifestly covariant BS model to the standard LFQM
using the Gaussian radial wave function.
The self-consistent covariant descriptions of the meson TFFs in the standard LFQM are given in this section.
In Sec.~\ref{sec:IV}, we present our numerical results for the
$(\pi^0,\eta,\eta')\to\gamma^*\gamma$ TFFs  and compare them with the available 
experimental data~~\cite{Babar09,Belle12,Babar11,CELLO91,CLEO98}. 
Summary and discussion follow in Sec.~\ref{sec:V}. 

\section{Model Calculation}
%\subsection{Two-point function: decay amplitude}
\label{sec:II}

\begin{figure*}
\centering
\includegraphics[height=3cm, width=10cm]{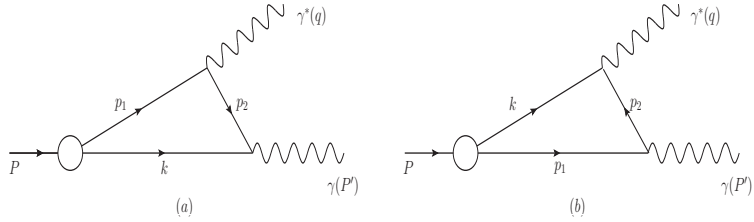}
\caption{\label{fig1}One-loop Feynman diagrams that contribute to $P\to\gamma^*\gamma$. }
\end{figure*}

The transition form factor $F_{P\gamma}$ for the $P\to\gamma^*\gamma$~($P=\pi^0, \eta$, and $\eta'$)
transition is defined from
the matrix element of electromagnetic current $\Gamma^\mu=\la\gamma(P-q)|J^\mu|P(P)\ra$
as follows:
%\be\label{Eq1}
$\Gamma^\mu = i e^2 F_{P\gamma}(Q^2)\ep^{\mu\nu\rho\sigma}P_\nu\ep_\rho q_\sigma$,
%\ee
where $P$ and $q$ are the momenta of the incident pseudoscalar meson and virtual photon,
respectively, and $\ep$ is the transverse polarization vector of the final (on-shell)
photon. This process is illustrated by the Feynman diagram in Fig.~\ref{fig1},
where Fig.~\ref{fig1}(a)~[1(b)] represents the amplitude
$\Gamma^\mu_{(a)}~[\Gamma^\mu_{(b)}]$ of the virtual photon being attached to the
quark~[antiquark] line. The total amplitude is then given by
$\Gamma^\mu_{\rm tot}=\Gamma^\mu_{(a)} + \Gamma^\mu_{(b)}$.
In the exactly solvable manifestly covariant BS model, the amplitude $\Gamma^\mu_{(a)}$
is given by the following momentum integral
\be\label{Eq2}
\Gamma^\mu_{(a)} = i e_Q e_{\bar Q} N_c
\int\frac{d^4k}{(2\pi)^4} \frac{{\rm Tr}\left[\gamma_5\left(\slash \!\!\!p_1 + m_Q \right)
 \gamma^\mu \left(\slash \!\!\!p_2 + m_Q \right)\slash \!\!\!\ep
 \left(-\slash \!\!\!k + m_Q \right) \right]}
{(p_{1}^2 -m_Q^2 +i\varepsilon)(p_{2}^2 -m_Q^2 +i\varepsilon)(k^2 -m_Q^2 +i\varepsilon)}H_0,
\ee
where $N_c$ is the number of colors and $e_{Q(\bar Q)}$ is the quark~(antiquark) electric charge
of mass $m_Q$ (=$m_{u(d)}, m_s$).
For the ${\bar q}q$ bound-state vertex function $H_0=H_0 (p^2_1, k^2)$ of the
meson, we simply take the constant parameter $g$ since the covariant loop
is regularized with this constant vertex in this model calculation.

Performing both manifestly covariant
calculation and the LF calculation of the amplitude  $I^{m_Q}_{(a)}(q^2)$ obtained from 
$\Gamma^\mu_{(a)} = ie_Q e_{\bar Q} I^{m_Q}_{(a)}(q^2)\ep^{\mu\nu\rho\sigma}P_\nu\ep_\rho q_\sigma$
in Eq.~(\ref{Eq2}), we explicitly  show the equivalence between the
two results $[I^{m_Q}_{(a)}]^{\rm Cov}$ and $[I^{m_Q}_{(a)}]^{\rm LF}$.
Especially for the LF calculation,
we take the reference frame where 
$P=(P^+, P^-, {\bf P}_\perp)=( P^+, M^2/P^+, 0)$
to investigate the LF zero-mode contribution. 
By the integration over $k^-$ in Eq.~(\ref{Eq2}) 
and using the plus component of the currents, we found that the LF zero-mode contribution
is absent and only the on-shell propagator contributes in the valence region.
The resulting LF amplitude $I^{m_Q}_{(a)}$ in this manifestly covariant model is given by
\be\label{Eq8}
 [I^{m_Q}_{(a)}]^{\rm LF}= \frac{N_c }{4\pi^3}\int^{1}_0
 \frac{dx}{x(1-x)} \int d^2{\bf k}_\perp
 \frac{m_Q}{M^{\prime 2}_0} \chi(x,{\bf k}_\perp),
\ee
where
\be\label{Eq9}
\chi(x,{\bf k}_\perp) = \frac{g}{x (M^2 -M^2_0)},
\ee
and $M^{(\prime)2}_0 = ({\bf k^{(\prime)}}^{2}_\perp + m^2_Q)/x (1-x)$
with ${\bf k'}_\perp = {\bf k}_\perp + (1-x){\bf q}_\perp$.
Likewise,  $[I^{m_Q}_{(b)}]^{\rm LF}$ corresponding to the second amplitude $\Gamma^+_{(b)}$ is obtained
as $[I^{m_Q}_{(b)}]^{\rm LF}$=$[I^{m_Q}_{(a)}]^{\rm LF} (x \to 1-x, {\bf q}_\perp \to -{\bf q}_\perp)$
but the two results are found to give the same numerical values. Thus, we obtain the total LF result  as
$I^{m_Q}_{\rm tot}= 2 [I^{m_Q}_{(a)}]^{\rm LF}$.

For $(\eta,\eta')\to\gamma^*\gamma$ transitions, we take into account the presence of two-nonstrange~($u$ and $d$) and 
strange~($s$)-components in the $\eta$ and $\eta'$ mesons as well as their mixing. Making use of the
$\eta-\eta'$ mixing scheme,
the flavor assignment of $\eta$ and $\eta'$ mesons in the quark-flavor basis $\eta_q=(u\bar{u}+d\bar{d})/\sqrt{2}$ and
 $\eta_s=s\bar{s}$ is given by~\cite{FKS}
  \be\label{Eq7a}
 \left( \begin{array}{cc}
 \eta\\
 \eta'
 \end{array}\,\right)
 =\left( \begin{array}{cc}
 \cos\phi\;\; -\sin\phi\\
 \sin\phi\;\;\;\;\;\cos\phi
 \end{array}\,\right)\left( \begin{array}{c}
 \eta_q\\
 \eta_s
 \end{array}\,\right).
 \ee

Therefore, we obtain the transition form factors
$F_{P\gamma}$ for $P\to\gamma^*\gamma ~(P=\pi^0, \eta, \eta')$ transitions
as follows~\cite{Jaus91}
\bea\label{Eq5}
F_{\pi\gamma}(q^2) &=& \frac{(e^2_u - e^2_d)}{\sqrt{2}} I^{m_{u(d)}}_{\rm tot},
\nonumber\\
F_{\eta\gamma} (q^2) &=& \cos\phi\; \frac{(e^2_u + e^2_d)}{\sqrt{2}} I^{m_{u(d)}}_{\rm tot}
- \sin\phi\; e^2_s  I^{m_s}_{\rm tot},
\nonumber\\
F_{\eta'\gamma} (q^2) &=& \sin\phi\;\frac{(e^2_u + e^2_d)}{\sqrt{2}} I^{m_{u(d)}}_{\rm tot}
+ \cos\phi\; e^2_s  I^{m_s}_{\rm tot},
\eea
where  $\phi$ is related with  
the mixing angle $\theta$ in the flavor SU(3) octet-singlet basis via
$\theta=\phi - {\rm arctan}\sqrt{2} \simeq \phi - 54.7^\circ$.

%\section{Application to Standard Light-Front Quark Model}
%\label{sec:III}
In the standard LFQM~\cite{CJ_99,Jaus91}, the
wave function of a ground state pseudoscalar meson as a $q\bar{q}$ bound state is given by
%
%\be\label{QM1}
$\Psi_{\lam{\bar\lam}}(x,{\bf k}_{\perp})
={\phi_R(x,{\bf k}_{\perp})\cal R}_{\lam{\bar\lam}}(x,{\bf k}_{\perp})$,
%\ee
%
where $\phi_R$ is the radial wave function and the
spin-orbit wave function ${\cal R}_{\lam{\bar\lam}}$
with the helicity $\lam({\bar\lam})$ of a quark~(antiquark)
is obtained by the interaction-independent Melosh transformation~\cite{Melosh}
from the ordinary spin-orbit wave function assigned by the quantum numbers $J^{PC}$.
The Gaussian wave function $\phi_R$ for $m_Q=m_{\bar Q}$
is given by
\be\label{QM2}
\phi_R(x,{\bf k}_{\perp}) = (4\pi^{3/4}/\beta^{3/2})
\sqrt{\partial k_z/\partial x} e^{m^2_Q/2\beta^2} e^{-M^2_0/8\beta^2},
\ee
where $\partial k_z/\partial x = M_0/4x(1-x)$ is the Jacobian of the variable transformation
$\{x,{\bf k}_\perp\}\to {\vec k}=({\bf k}_\perp, k_z)$
and $\beta$ is the variational parameter
fixed by the analysis of meson mass spectra~\cite{CJ_99}.
In our previous analysis of the twist-2 and twist-3 DAs of
pseudoscalar and vector mesons~\cite{TWV,TWPS} and the pion electromagnetic form factor~\cite{TWPS},
we have shown that standard LF (SLF) results of the LFQM is obtained by the
replacement of the LF vertex function $\chi$ in the BS model with the Gaussian wave function
$\phi_R$ as follows  [see, e.g., Eq. (35) in~\cite{TWPS}]
\be\label{QM7}
 \sqrt{2N_c} \frac{ \chi(x,{\bf k}_\perp) } {1-x}
 \to \frac{\phi_R (x,{\bf k}_\perp) }
 {\sqrt{{\bf k}^2_\perp + m_Q^2}}, \; M \to M_0,
 \ee
where $M\to M_0$ implies that the physical mass $M$ included in the integrand of BS
amplitude has to be replaced with the invariant
mass $M_0$ since the SLF results in the LFQM
are obtained from the requirement of all constituents being on their respective mass-shell.
The correspondence in Eq.~(\ref{QM7}) is valid again in this analysis of  $P \to \gamma^*\gamma$ transition.
Applying the correspondence given by Eq.~(\ref{QM7}) to Eq.~(\ref{Eq8}),
we obtain the corresponding SLF result $[I^{m_Q}_{(a)}]^{\rm SLF}$ in the LFQM and the total result 
$[I^{m_Q}_{\rm tot}]^{\rm LFQM} = 2 [I^{m_Q}_{(a)}]^{\rm SLF}$ is given by
\be\label{QM8}
[I^{m_Q}_{\rm tot}]^{\rm LFQM} = \frac{\sqrt{2 N_c}}{4\pi^3}\int^{1}_0
 \frac{dx}{x(1-x)} \int d^2{\bf k}_\perp
 \frac{\phi_R(x,{\bf k}_{\perp})}{\sqrt{{\bf k}^2_\perp + m^2_Q}}
 \frac{(1-x) m_Q}{M^{\prime 2}_0}.
\ee
Our LFQM predictions of the TFFs for $P\to\gamma^*\gamma$ is then obtained by substituting
$[I^{m_Q}_{\rm tot}]^{\rm LFQM}$ into Eq.~(\ref{Eq5}).

\section{Numerical Results}
\label{sec:IV}
In our numerical calculations, we use the constituent quark masses $(m_{q}, m_s)=(0.22, 0.45)$ GeV 
and the gaussian parameters $(\beta_{q{\bar q}}, \beta_{s{\bar s}})=(0.3695,0.4128)$ GeV 
($q=u$ and $d$), which were obtained from the calculation of meson mass spectra using
the variational principle in our LFQM~\cite{CJ_99,CJ_DA}.
While the quadratic (linear) Gell-Mann-Okubo mass formula prefers
$\theta\simeq -10^\circ, \phi\simeq 44.7^\circ$
($\theta\simeq -23^\circ, \phi\simeq 31.7^\circ$),
the KLOE Collaboration~\cite{KLOE} extracted $\phi=(41.5\pm 0.3_{\rm stat}\pm 0.7_{\rm syst}\pm 0.6_{\rm th})^{\circ}$
by measuring the ratio ${\rm BR}(\phi\to\eta'\gamma)/{\rm BR}(\phi\to\eta\gamma)$ and
%The measured values are $\phi=(39.7\pm 0.7)^{\circ}$ and
%$(41.5\pm 0.3_{\rm stat}\pm 0.7_{\rm syst}\pm 0.6_{\rm th})^{\circ}$
%with and without the gluonium content for $\eta'$, respectively.
RBC-UKQCD Collaboration~\cite{RBC} obtained $\phi=40.6(2.8)^\circ$.
We thus use $\phi=37^\circ\sim 42^\circ$ to check the sensitivity of our LFQM
since the mixing angle for $\eta-\eta'$ is still not yet settled as a fixed value.

\begin{figure*}
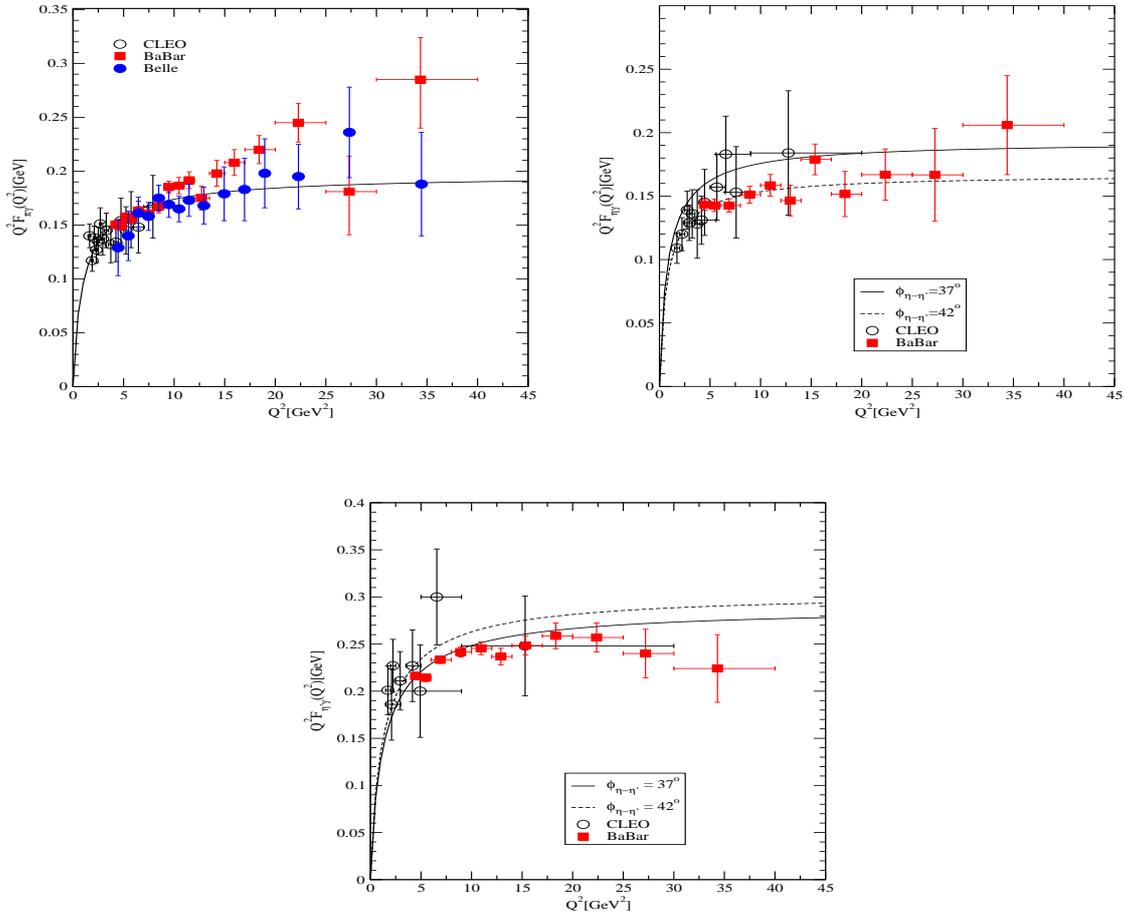

\vspace{0.7cm}
\centering
\includegraphics[height=5.5cm, width=7cm]{Fig2.eps}
\hspace{0.5cm}
%\\
\includegraphics[height=5.5cm, width=7cm]{Fig3.eps}
\vspace{1cm}
\\
\includegraphics[height=5.5cm, width=7cm]{Fig4.eps}
\caption{\label{fig3} The transition form factors $Q^2 F_{\pi\gamma}(Q^2)$, 
$Q^2 F_{\eta\gamma}(Q^2)$,  and $Q^2 F_{\eta'\gamma}(Q^2)$
up to $Q^2=45$ GeV$^2$. 
For $Q^2 F_{(\eta,\eta')\gamma}(Q^2)$ case, the solid and dashed lines are results obtained from $\eta-\eta'$ mixing angles with
$\phi_{\eta-\eta'}=37^\circ$ and $42^\circ$, respectively.
The data are taken from~\cite{Babar09,Belle12,Babar11,CELLO91,CLEO98}. }
\end{figure*}

In Fig.~\ref{fig3}, we show the transition form factors $Q^2 F_{\pi\gamma}(Q^2)$, 
$Q^2 F_{\eta\gamma}(Q^2)$,  and $Q^2 F_{\eta'\gamma}(Q^2)$ up to $Q^2=45$ GeV$^2$. 
For $Q^2 F_{(\eta,\eta')\gamma}(Q^2)$ case, the solid and dashed lines are results obtained from $\eta-\eta'$ mixing angles with
$\phi_{\eta-\eta'}=37^\circ$ and $42^\circ$, respectively. The data are taken from~\cite{Babar09,Belle12,Babar11,CELLO91,CLEO98}.
For $Q^2 F_{\eta\gamma}(Q^2)$, we obtain the asymptotic result as
$\lim_{Q\to\infty}Q^2F_{\pi\gamma}(Q^2)\simeq 0.195$ GeV, which is
consistent with the asymptotic limit set by perturbative QCD:
$Q^2F_{\pi\gamma}(Q^2)=\sqrt{2}f_\pi\simeq 0.185$ GeV~\cite{BL80}.
We also note that our LFQM result for $Q^2 F_{\pi\gamma}(Q^2)$  is in good agreement with the recent data from the Belle experiment~\cite{Belle12}
showing the asymptotic behavior
for the region $10 \leq Q^2\leq 45$ [GeV$^2$]
but disagree with the BaBar data~\cite{Babar09} showing the rapid growth for this $Q^2$ regime.
For $Q^2 F_{\eta\gamma}(Q^2)$  and $Q^2 F_{\eta'\gamma}(Q^2)$ TFFs, our predictions $F_{\eta\gamma}(Q^2)$ 
using the mixing angle $\phi=37^\circ$ show slightly better agreement with the data compared to the results obtained
from the mixing angle $\phi=42^\circ$. As in the case of $Q^2F_{\pi\gamma}(Q^2)$, 
the TFFs $Q^2F_{\eta\gamma}(Q^2)$  and $Q^2F_{\eta'\gamma}(Q^2)$ show asymptotic behavior for
high $Q^2$  region. The asymptotic values obtained in the spacelike region are obtained as follows:
$\lim_{Q^2\to\infty}Q^2 F_{\eta\gamma}(Q^2)\simeq 0.192 ~(0.167)\;{\rm GeV}\; {\rm for}\;\phi=37^\circ~(42^\circ)$ and 
$\lim_{Q^2\to\infty}Q^2 F_{\eta'\gamma}(Q^2)\simeq 0.286 (0.302)\;{\rm GeV}\; {\rm for}\;\phi=37^\circ(42^\circ)$, respectively.

\section{Summary and Discussion}
\label{sec:V}
In this work, we investigated $(\pi^0,\eta,\eta')\to\gamma^*\gamma$ TFFs using the standard LF (SLF) approach 
within the phenomenologically accessible, realistic LFQM~\cite{CJ_99,CJ_DA}.  As the SLF approach within the LFQM by itself 
is not amenable to determine the zero-mode contribution, we utilized the covariant BS model to check the existence (or absence) 
of the zero mode as we discussed in~\cite{TWV,TWPS}.
Performing a LF calculation in the covariant BS model,
we found that the TFF using the plus component of the currents is immune to the zero-mode.
We then linked the covariant BS model to the standard LFQM following the same correspondence relation Eq.~(\ref{QM7}) between
the two  that we found in our previous analysis of two-point and three-point functions for pseudoscalar and vector
meson~\cite{TWV,TWPS}. This link allows us to effectively substitute the LF vertex function in the covariant BS model with the more phenomenologically accessible
Gaussian wave function provided by the LFQM analysis of meson mass~\cite{CJ_99,CJ_DA}.

For the $\pi\to\gamma^*\gamma$ transition, our numerical result of $Q^2F_{\pi\gamma}(Q^2)$ does not show any steep rising  behavior for high $Q^2$ region
as measured from the BaBar Collaboration~\cite{Babar09} but
shows scaling behavior for high $Q^2$ consistent with the pQCD prediction.
This may be ascribed to the fact that our twist-2 DA~\cite{CJ_DA,TWPS} is highly
suppressed at the end points ($x=0,1$) unlike the flat DA~\cite{Ra09,MP09} showing the
enhancement at the end points.  We should note that our results for the twist-2 pion DA and the pion-photon
transition form factor are very similar to those obtained by the authors~\cite{Nico2}.
For the $(\eta,\eta')\to\gamma^*\gamma$ transitions, we use the $\eta-\eta'$ mixing angles $\phi=[37^\circ,42^\circ]$
in the quark-flavor basis to check the sensitivity of our LFQM. Comparing the experimental data for  
$Q^2 F_{(\eta,\eta')\gamma}(Q^2)$, our optimum value of the $\eta-\eta'$ mixing angle seems to  be $\phi\simeq 37^\circ$.
However, more experimental data in the asymptotic region are needed to pin down more accurate $\eta-\eta'$ mixing angle.
Our results of  $Q^2 F_{(\eta,\eta')\gamma}(Q^2)$ show again scaling behavior for  high  $Q^2$ consistent with the pQCD prediction.

%\begin{acknowledgements}
%\end{acknowledgements}
% BibTeX users please use
%\bibliographystyle{spbasic}

\end{document}